%% file: paper.tex
\documentclass[12pt]{article}
\usepackage{epsfig}
\usepackage{amsmath}
\usepackage{hhline}
\usepackage{amssymb}
\usepackage{times}
\usepackage{cite}
\usepackage{rotating}

\newlength{\dinwidth}
\newlength{\dinmargin}
\begin{document}  
\newcommand{\kev}{\,\mbox{keV}}
\newcommand{\mev}{\,\mbox{MeV}}
\newcommand{\gev}{\,\mbox{GeV}}
\newcommand{\mub}{\,\mu\mbox{b}}
\newcommand{\mrad}{\,\mbox{mrad}}
\newcommand{\meter}{\,\mbox{m}}
\newcommand{\mum}{\mu\mbox{m}}
\newcommand{\cm}{\,\mbox{cm}}
\newcommand{\mm}{\,\mbox{mm}}
\newcommand{\mb}{\,\mbox{mb}}
\newcommand{\nb}{\,\mbox{nb}}
\newcommand{\der}{\mbox{d}}
\newcommand{\pttrk}{p_T^{\mathrm{trk}}}
\newcommand{\ptproton}{p_T}
\newcommand{\zvtx}{z_{\mathrm{Vtx}}}
\newcommand{\invpb}{\,\mbox{pb}^{-1}}
\newcommand{\ftwolpthree}{F_2^{LP(3)}}
\newcommand{\ftwovdm}{F_2^{VDM}}
\newcommand{\ftwopart}{F_2^{part}}
\newcommand{\fraclp}{{\mathrm{f}}_{\mathrm{LP}}}
\newcommand{\cvm}{C_{VM}}
\newcommand{\qsqvm}{Q^2_{VM}}
\newcommand{\cvmlp}{C_{VM}^{LP}}
\newcommand{\pom}{I\!\!P}
\newcommand{\reg}{I\!\!R}
\newcommand{\slowpi}{\pi_{\mathit{slow}}}
\newcommand{\fiidiii}{F_2^{D(3)}}
\newcommand{\fiidiiiarg}{\fiidiii\,(\beta,\,Q^2,\,x)}
\newcommand{\n}{1.19\pm 0.06 (stat.) \pm0.07 (syst.)}
\newcommand{\nz}{1.30\pm 0.08 (stat.)^{+0.08}_{-0.14} (syst.)}
\newcommand{\fiidiiiful}{F_2^{D(4)}\,(\beta,\,Q^2,\,x,\,t)}
\newcommand{\fiipom}{\tilde F_2^D}
\newcommand{\ALPHA}{1.10\pm0.03 (stat.) \pm0.04 (syst.)}
\newcommand{\ALPHAZ}{1.15\pm0.04 (stat.)^{+0.04}_{-0.07} (syst.)}
\newcommand{\fiipomarg}{\fiipom\,(\beta,\,Q^2)}
\newcommand{\pomflux}{f_{\pom / p}}
\newcommand{\nxpom}{1.19\pm 0.06 (stat.) \pm0.07 (syst.)}
\newcommand {\gapprox}
   {\raisebox{-0.7ex}{$\stackrel {\textstyle>}{\sim}$}}
\newcommand {\lapprox}
   {\raisebox{-0.7ex}{$\stackrel {\textstyle<}{\sim}$}}
\def\gsim{\,\lower.25ex\hbox{$\scriptstyle\sim$}\kern-1.30ex%
\raise 0.55ex\hbox{$\scriptstyle >$}\,}
\def\lsim{\,\lower.25ex\hbox{$\scriptstyle\sim$}\kern-1.30ex%
\raise 0.55ex\hbox{$\scriptstyle <$}\,}
\newcommand{\pomfluxarg}{f_{\pom / p}\,(x_\pom)}
\newcommand{\dsf}{\mbox{$F_2^{D(3)}$}}
\newcommand{\dsfva}{\mbox{$F_2^{D(3)}(\beta,Q^2,x_{I\!\!P})$}}
\newcommand{\dsfvb}{\mbox{$F_2^{D(3)}(\beta,Q^2,x)$}}
\newcommand{\dsfpom}{$F_2^{I\!\!P}$}
\newcommand{\gap}{\stackrel{>}{\sim}}
\newcommand{\lap}{\stackrel{<}{\sim}}
\newcommand{\fem}{$F_2^{em}$}
\newcommand{\tsnmp}{$\tilde{\sigma}_{NC}(e^{\mp})$}
\newcommand{\tsnm}{$\tilde{\sigma}_{NC}(e^-)$}
\newcommand{\tsnp}{$\tilde{\sigma}_{NC}(e^+)$}
\newcommand{\st}{$\star$}
\newcommand{\sst}{$\star \star$}
\newcommand{\ssst}{$\star \star \star$}
\newcommand{\sssst}{$\star \star \star \star$}
\newcommand{\tw}{\theta_W}
\newcommand{\sw}{\sin{\theta_W}}
\newcommand{\cw}{\cos{\theta_W}}
\newcommand{\sww}{\sin^2{\theta_W}}
\newcommand{\cww}{\cos^2{\theta_W}}
\newcommand{\trm}{m_{\perp}}
\newcommand{\trp}{p_{\perp}}
\newcommand{\trmm}{m_{\perp}^2}
\newcommand{\trpp}{p_{\perp}^2}
\newcommand{\alp}{\alpha_s}

\newcommand{\alps}{\alpha_s}
\newcommand{\sqrts}{$\sqrt{s}$}
\newcommand{\LO}{$O(\alpha_s^0)$}
\newcommand{\Oa}{$O(\alpha_s)$}
\newcommand{\Oaa}{$O(\alpha_s^2)$}
\newcommand{\PT}{p_{\perp}}
\newcommand{\JPSI}{J/\psi}
\newcommand{\sh}{\hat{s}}
\newcommand{\uh}{\hat{u}}
\newcommand{\MP}{m_{J/\psi}}
\newcommand{\PO}{I\!\!P}
\newcommand{\xbj}{x}
\newcommand{\xpom}{x_{\PO}}
\newcommand{\ttbs}{\char'134}
\newcommand{\xpomlo}{3\times10^{-4}}  
\newcommand{\xpomup}{0.05}  
\newcommand{\dgr}{^\circ}
\newcommand{\pbarnt}{\,\mbox{{\rm pb$^{-1}$}}}
\newcommand{\WBoson}{\mbox{$W$}}
\newcommand{\fbarn}{\,\mbox{{\rm fb}}}
\newcommand{\fbarnt}{\,\mbox{{\rm fb$^{-1}$}}}
\newcommand{\stat}{\,{\rm (stat) }}
\newcommand{\syst}{\,{\rm (syst) }}
%
%
\newcommand{\qsq}{\ensuremath{Q^2} }
\newcommand{\gevsq}{\ensuremath{\mathrm{GeV}^2} }
\newcommand{\et}{\ensuremath{E_t^*} }
\newcommand{\rap}{\ensuremath{\eta^*} }
\newcommand{\gp}{\ensuremath{\gamma^*}p }
\newcommand{\dsiget}{\ensuremath{{\rm d}\sigma_{ep}/{\rm d}E_t^*} }
\newcommand{\dsigrap}{\ensuremath{{\rm d}\sigma_{ep}/{\rm d}\eta^*} }
\def\Journal#1#2#3#4{{#1} {\bf #2} (#3) #4}
\def\NCA{\em Nuovo Cimento}
\def\NIM{\em Nucl. Instrum. Methods}
\def\NIMA{{\em Nucl. Instrum. Methods} {\bf A}}
\def\NPB{{\em Nucl. Phys.}   {\bf B}}
\def\PLB{{\em Phys. Lett.}   {\bf B}}
\def\PRL{\em Phys. Rev. Lett.}
\def\PRD{{\em Phys. Rev.}    {\bf D}}
\def\ZPC{{\em Z. Phys.}      {\bf C}}
\def\EJC{{\em Eur. Phys. J.} {\bf C}}
\def\CPC{\em Comp. Phys. Commun.}

\begin{titlepage}

\begin{flushleft}
DESY 01--062 \hfill ISSN 0418--9833 \\
May 2001
\end{flushleft}

\vspace{2.2cm}

\begin{center}
\begin{Large}

{\bf Photoproduction with a Leading Proton at HERA}

\vspace{1.8cm}

H1 Collaboration

\end{Large}
\end{center}

\vspace{1.3cm}

\begin{abstract}
The total cross section for the photoproduction process with a 
leading proton in the final state has been measured at 
$\gamma p$~centre-of-mass energies~$W$ of
$91$, $181$ and $231\gev$. 
The measured cross sections apply to the kinematic range with    
the transverse momentum of the scattered proton restricted to
$\ptproton \leq 0.2 \gev$ and $0.68 \leq z \leq 0.88$,
where $z = E_p'/E_p$ is the 
scattered proton energy normalised to the beam energy. 
The cross section 
$\der\sigma_{\gamma p \to Xp'}(W,z)/\der z$ is observed 
to be independent of~$W$ and $z$ within the measurement errors and
amounts to 
$(8.05 \pm 0.06 \stat \pm 0.89 \syst)\mub$ on average.
The data are well described by a Triple Regge model in which
the process is mediated by a mixture of
exchanges with an effective Regge trajectory of intercept
$\alpha_i(0)=0.33 \pm 0.04 \stat \pm 0.04\syst$.
The total cross section for the interaction of 
the photon with this mixture ($\gamma \alpha_i \rightarrow X$)
can be described by an effective trajectory of intercept
$\alpha_k(0)=0.99 \pm 0.01\stat  \pm 0.05 \syst$.
Predictions based on previous
triple Regge analyses of $pp \rightarrow pX$ data assuming vertex
factorisation are broadly consistent with the $\gamma p$ data.
The measured cross sections are compared with 
deep inelastic scattering leading proton data 
in the same region of $z$ and $\ptproton$ for photon virtuality
$Q^2 > 2.5 \ {\rm GeV^2}$.
The ratio of the cross section for leading proton
production to the total cross section is found to rise
with $Q^2$.
\end{abstract}

\vspace{1.5cm}

\begin{center}
To be submitted to \NPB
\end{center}

\end{titlepage}

%
%

\begin{flushleft}
  \input{h1auts}
\end{flushleft}

\newpage

\section{Introduction}
Over the years there has been considerable interest in ``diffractive'' 
dissociation
processes in strong interactions. At $pp$~collider experiments, the
reaction $pp \to Xp'$ has been studied in 
detail~\cite{Goulianos,diffrinpbarp}, whilst at
HERA the process $\gamma p \to Xp'$ has also been closely
investigated for both real and virtual 
photons~\cite{lbproduction,H1dd,ZEUSdd,H1f2d3,ZEUSf2d3,zeusf2d4}.
These processes are often interpreted using Regge phenomenology in terms
of the exchange of colourless objects between the colliding particles.
Such scattering is distinguished in collider experiments by the observation of
rapidity gaps between the beam direction and the produced
hadrons,~$X$. A complementary way to study these
processes is by direct detection of the leading protons from the reactions. 

At HERA, the nature of the exchange has been found to depend on the
fraction of the beam longitudinal momentum retained by the final
state proton,~$z$\footnote{In some publications, 
$\xpom=1-z$ is used instead.}.
When $z \to 1$, the cross section is well described by
pomeron ($\pom$) exchange in both photoproduction~\cite{H1dd,ZEUSdd} and
Deep Inelastic Scattering (DIS)~\cite{H1f2d3,ZEUSf2d3}.
At $z\sim 0.95$, additional 
meson trajectories associated with $\rho$, $\omega$,
$f_2$ or $a_2$, here collectively
denoted~$\reg$, have been found to 
contribute~\cite{H1dd,H1f2d3}.
In a previous publication~\cite{lbproduction}
DIS data with a leading proton in the kinematic range $0.73\leq z \leq 0.88$
were shown to be well described by a Regge model in which the
virtual photon interacts
with a mixture of exchanges dominated by $\reg$ and $\pi$.
In this
paper, we present measurements using the H1~Forward Proton
Spectro\-meter (FPS) to reconstruct the proton momentum
directly in a similar region of $z$
to~\cite{lbproduction}, but in the photoproduction limit. These
measurements extend those of~\cite{H1dd} to lower values of~$z$.

\begin{figure}[ht]
\begin{center}
  \setlength{\unitlength}{1mm}
  \epsfig{file=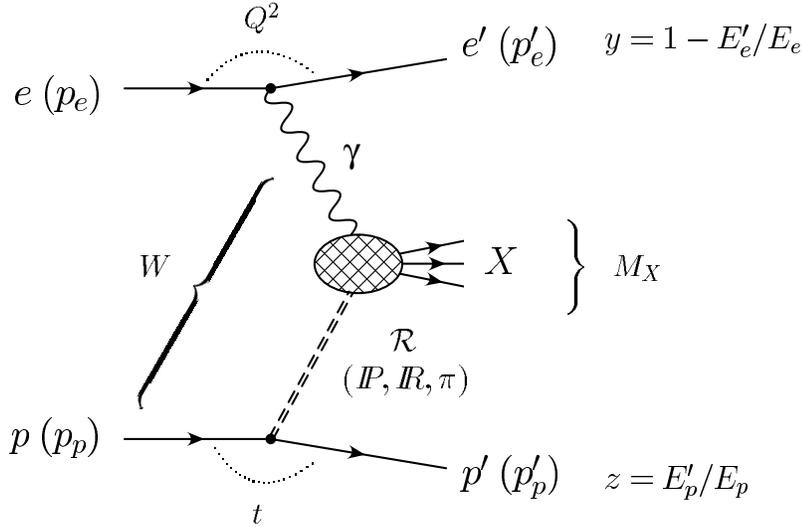}
  \caption[]{
  \sl The leading proton process $e p \to e'p'X$ in a Regge exchange model. 
  At the proton vertex,
  $z=E_p'/E_p$ is the energy fraction retained by the proton.
  \label{picfeyndiag}
  }
\end{center}
\end{figure}

The semi-inclusive scattering reaction $ep\to e' X p'$ is sketched in
figure~\ref{picfeyndiag}. The measured process described in this paper is
$\gamma p \to X p'$ for quasi-real photons, in which
the photon emitted by the incoming electron reacts with the proton
to produce a system of mass~$M_X$.
The squared four-momentum transfer at the proton vertex is denoted by~$t$.
The proton remains intact
leaving the interaction with an energy fraction~$z=E_p'/E_p$, 
$E_p'$ and $E_p$ being the energies of the outgoing and
incoming protons, respectively. 
For the elastic process shown in figure~\ref{picfeyndiag}, 
the mass $M_X$ is equal to $W\sqrt{1-z}$,  
where $W$~is the centre-of-mass energy of the photon-proton system.
In a Regge interpretation, the reaction proceeds via the exchange of
an object~${\cal R}$.

The $W$ and $z$ dependences of the data are used to investigate 
the contributions
of different exchanges away from the pomeron dominated region.
Comparison is made with measurements of the
process $pp \rightarrow X p^\prime$ 
using a model based on Regge phenomenology.
Together with the data from DIS in~\cite{lbproduction}
the photon virtuality,~$Q^2$, dependence of the 
semi-inclusive scattering process is investigated.
This allows one to study how the dissociative process varies as the
incoming particle changes from a point-like object at high~$Q^2$
to a hadron-like object in the photoproduction region at~$Q^2=0$.
This transition has
been studied in detail in inclusive scattering
$e p \rightarrow e'X$~\cite{h1f2fits,zeuslowq2},
but no previous information exists for semi-inclusive
processes at the values of~$z$ studied here. 

\section{Experimental Method}
\subsection{The H1 Detector}
The H1 detector is described in detail elsewhere~\cite{theh1detathera}.
We use
a coordinate system originating at the interaction point with
the positive $z$~axis along the proton beam direction.
The selection of the final state of the reaction under study is performed
by identifying the scattered proton, the scattered electron and at
least one charged track in the central region of the H1~detector.
The central tracking
chambers and the central proportional chambers, located within  
the $1.15$~Tesla field of the H1 solenoid, trigger on and measure the momentum
of tracks, the resolution being 
$\Delta(p_t)/p_t^2 \sim 10^{-2} \gev^{-1}$
for $p_t > 0.5\gev$.
Small angle electron calorimeters and a photon detector are
used to determine the luminosity via the Bethe-Heitler process
$ep \to e'p'\gamma$, to tag photoproduction
events and to suppress Bethe-Heitler background in the 
measurement.   
By selecting an electron candidate in the electron calorimeters,
the acceptance is restricted to values of the photon virtuality
~$Q^2<0.01\gev^2$.

The Forward Proton Spectrometer (FPS)~\cite{lbproduction,fpsnimpaper}
is used to measure the energy and scattering angles at the interaction
point of the outgoing proton.  Protons scattered at small angles to
the incident proton direction are deflected by the magnets of the beam
optics into a system of detectors placed close to the proton beam in
moveable ``Roman pot'' housing stations, approaching the beam from
above.  The stations used in this analysis are positioned at distances
of $81$ and $90$~m downstream of the interaction point and are moved into
position once stable beam conditions are established.
The pattern of hits from many events observed in the position
detectors is analyzed to determine spatial offsets and tilts of the
proton beam at the interaction point.  For each event the outgoing
proton energy and scattering angles are then obtained from the
reconstructed track positions and the measured interaction point using
the transfer functions derived from the beam
optics~\cite{bennosthesis}.  The energy resolution deteriorates with
energy and is better than $8\gev$.  The absolute energy scale
uncertainty is estimated to $\pm 10\gev$, as has been inferred from a
comparison of a diffractive model with the data and from a study of a
small number of events at large $Q^2$, where the kinematics are
determined independently in the main
detector~\cite{fpsnimpaper,bennosthesis}.  The mean error on the angle
measurement is $5$~$\mu$rad in the $\theta_x$ projection and varies
with increasing beam energy from $5$~$\mu$rad to $100$~$\mu$rad in the
$\theta_y$ projection.
With a primary proton beam of $820\gev$, scattered proton energies in
the range $500\gev < E_p'< 780\gev$ are accepted by the spectrometer. 
For a proton which passes through both stations, the average overall
track reconstruction efficiency is approximately $50$\%.

\subsection{Data Selection}
The analysis is based on a data set from the 1996 running 
period, when HERA collided $820\gev$ protons with
$27.5\gev$ positrons, resulting in an $ep$ centre-of-mass
energy $\sqrt s = 300\gev$.
Further details of this analysis can be found
in~\cite{hannasthesiscarstensthesis}.
Data were analysed where the FPS was in a stable 
position close to the circulating beam and all relevant components of 
the H1 detector
were fully operational. The corresponding integrated
luminosity amounts to $3.3\invpb$.
In order to study the reaction $\gamma p \to Xp'$, events were
selected with a reconstructed track in the FPS,
a positron candidate in one
of the small-angle electron detectors of the luminosity system
and at least one reconstructed track in the central H1 detector. 
At the trigger level, a signal in each of the components was required.
Events with energy deposition in the photon detector of the 
luminosity system were rejected,
since this indicates the presence of an event from the Bethe-Heitler 
process in the same
bunch crossing or initial or final state radiation.
 
At least one track in the H1 detector was required to have a
transverse momentum above $0.5\gev$   
and a polar angle~$\theta$ in the range
of $20^\circ < \theta < 160^\circ$ in order to guarantee a good
trigger and reconstruction efficiency. 
In order to reject background from interactions of the beam with the
residual gas in the beam pipe,
the $z$ coordinate of the event vertex was restricted to 
$|\zvtx| <35 \cm$.
Further background rejection conditions were imposed, which require  the 
timing of the event to match the bunch crossing time.

The scattered proton energy range was restricted to $540 \gev < E_p' <
740 \gev$ and the proton transverse momentum range to $0 < p_T \leq
0.2\gev$, where the FPS acceptance is well understood.  The data were
grouped into five $40\gev$ wide intervals in $E_p'$.  The scattered
electron energies were measured in two calorimeters allowing the data
to be divided into three ranges of~$W$.  The average values of~$W$ in
each of these ranges were $91\gev$, $187\gev$ and $231\gev$.  This
$W$~range together with the measured $z$~range corresponds to values
of $M_X$ between $32\gev$ and $130\gev$, assuming the scattering is
elastic at the proton vertex.  The total number of events selected
was~$23072$.

\subsection{Cross Section Measurement}
\label{seccrosssection}
To obtain the differential cross section 
$\der^3\sigma_{ep \to e'Xp'}/\der y \der Q^2 \der z$,
where $y = W^2 / s$ and where the scattered proton $p_T$ is integrated
over the range $0 < p_T < 0.2\gev$,
the data are corrected for acceptances and efficiencies.
The PHOJET~\cite{phojetman} Monte Carlo programme,
a general photoproduction model, 
was used to correct for central track reconstruction efficiencies.
The acceptance of the FPS was studied using both
PHOJET and POMPYT~\cite{pompytman}, which is a Monte Carlo model
for diffractive hard scattering, also capable of describing
reactions mediated by pion exchange.
Corrections are made for the FPS trigger efficiency,
the FPS track reconstruction efficiency, 
the track multiplicity dependent trigger efficiency of the central
tracker
and the electron tagger acceptances. The latter
includes a correction for cases where events are lost due to an 
energy deposition in the photon detector by 
an overlaid Bethe-Heitler event. Bin by bin
factors are used to correct for
track reconstruction inefficiency in the H1 tracking
chambers, to compensate for the
migrations between proton energy intervals and 
to take the limited FPS acceptance into account. 
The values for the
migration corrections vary between $0.91$ and $1.13$ 
for the intervals defined in section~\ref{secresults}. 
The correction factor for the fraction of events lying outside the FPS
acceptance region ranges between $0.27$ to $0.90$ with an average
value of about $0.70$.

The differential cross section 
$\der\sigma_{\gamma p \to Xp'} (W,z) /\der z$ is determined 
using the relation
\begin{equation}
\frac{\der^3\sigma_{e p \to eXp'}(W,Q^2,z)}{\der y \der Q^2 \der z}
= {\cal F}_{\gamma /e}(y,Q^2)
\frac{\der\sigma_{\gamma p \to Xp'}(W,z)}{\der z},
\end{equation}
where ${\cal F}_{\gamma /e}(y,Q^2)$ is the photon flux
in the Equivalent Photon Approximation~\cite{equivphotonapprox}.
The effects of initial and final state radiation from the
electron are suppressed in this analysis by the veto of events
with energy deposition in the small angle photon detector.
Radiative corrections are thus expected to be small~\cite{h1gammap} and  
have been neglected throughout the analysis.

The measured cross sections are defined solely in terms of the specified 
ranges in the scattered proton transverse momentum, $W$~and~$z$. 
No subtractions have been made for the 
contribution of protons from the decay of baryon resonances or proton
dissociation processes. The 
major source, the $\Delta(1232)$, has been estimated to contribute 
at the $10\%$~level to the total cross section for leading proton production
in the range $0.6 < z < 0.9$ in DIS~\cite{Szczurek},
concentrated at low $z$. Assuming a contribution of
similar magnitude in photoproduction, this will be suppressed to the
percent level by the restriction of this measurement 
to proton transverse momenta~$\ptproton < 0.2\gev$ 
and $z>0.66$.

\subsection{Background}
The following sources of background were studied:
\begin{itemize}
\item Protons in the FPS originating from
interactions of the proton beam with the beam pipe wall or residual
gas: 
the background actually entering the sample where a trajectory is 
indistinguishable from
that of a proton which was scattered in an $ep$~process at
the interaction point has been found to 
be much less than~1$\%$ from studies of non-interacting
proton bunches. This background has been neglected.

\item Tracks in the central H1 detector produced
by beam gas interactions: this background has been estimated 
using the distribution of $\zvtx$. This is approximately
distributed as a Gaussian around a mean value
close to the nominal interaction point for $ep$~interactions
while beam gas interactions have a more uniform distribution in~$\zvtx$.
The fraction of background events entering the
data sample, where $|\zvtx|<35\cm$ is required,
can be estimated from the tails
of the $\zvtx$ distribution and has been
determined to vary between $2.2\%$ (lowest $y$ interval)
and $3.4\%$ (highest $y$ interval). 
This has been subtracted.

\item Bethe-Heitler events $ep\to e'p'\gamma$: these are not a 
significant source of background unless they overlap
with a photoproduction event. The acceptance
of the photon detector of the luminosity system
for the photons is large.
With the cut imposed on the maximum energy deposited 
in the photon detector, 
this background has been suppressed to the level of $0.25\%$,
determined statistically from the probability of
random overlap. This has been subtracted. 
\end{itemize}

\subsection{Systematic errors}
The systematic uncertainties in the measurement can be grouped
into those depending on~$y$, those related to the proton energy 
(or~$z$) interval and normalisation errors.
\paragraph{Errors depending on \boldmath $y$.}
The following sources contribute to this type of
error: 
errors on the acceptance of the electron taggers;
errors on the selection efficiency for tracks in the H1
tracking chambers;
migration uncertainties between adjacent $y$~intervals and
errors on the proton beam induced background 
in the central region of the H1 detector. 
The resulting uncertainties were found to vary between $6.5$ and $8.5\%$.
\paragraph{Errors depending on \boldmath $z$.}
The largest uncertainty arises from the acceptance
correction for migrations
about the measurement limit at the proton transverse
momentum of $p_T=0.2\gev$.
The uncertainty was evaluated using the PHOJET
and POMPYT Monte Carlo generators. 
Uncertainties in the FPS calibration constants lead to systematic errors 
on the migration corrections between proton energy
bins. This effect has also been studied using the PHOJET and POMPYT Monte
Carlo models.
The $z$~dependent errors were found to vary 
between~$2.6$ and~$14.1\%$.
\paragraph{Normalisation errors.}
These are dominated by overall FPS uncertainties such as alignment errors and
uncertainties on the hodoscope efficiencies.
The uncertainty in the luminosity measurement, in
the vertex cut efficiency and in the
positron beam related background are also included. 
This error amounts to~$5.3\%$.

\section{Results}
\subsection{The differential cross section \boldmath
  $\der\sigma_{\gamma p \to Xp'}(W,z)/\der z$}
\label{secresults}
\begin{figure}[ht]
    \begin{center}
        \setlength{\unitlength}{1mm}
        \epsfig{file=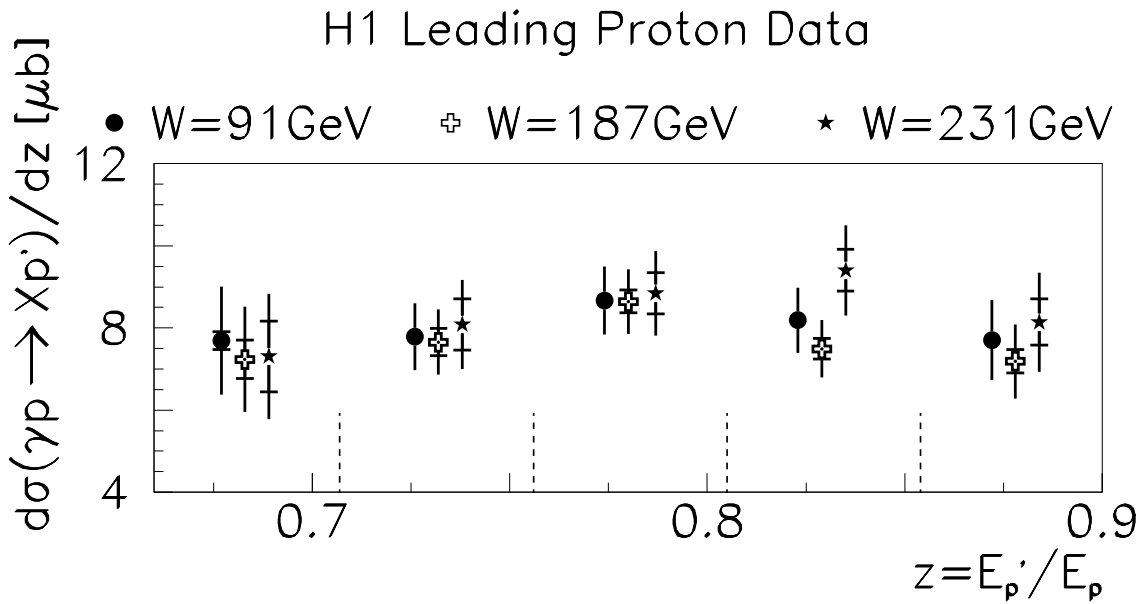,width=15cm}
        \caption[]{
  \sl The cross section $\der \sigma_{\gamma p\to Xp'}/\der z$ as a
     function of $z$ for three values of $W$.  
     The inner error bar is the statistical and the outer is the
     total error (statistical and systematic error added in
     quadrature). The dashed lines show the limits of the $z$ intervals
     from which the differential cross sections are obtained.
     The data points for
     $W=91$ and $W=231\gev$ are displaced in~$z$ for 
     visibility. \label{sigmaplot}
        } 
    \end{center}
\end{figure}

The cross section $\der\sigma_{\gamma p \to Xp'}(W,z)/\der z$
for $W=91$, $187$ and $231\gev$ and five values of~$z$ in the range 
$0.68 \leq z \leq 0.88$ is shown in figure~\ref{sigmaplot}. The
values are listed in Table~\ref{tabxsect}. 
It is observed that for all $z$ and $W$,
the measured cross sections are compatible 
with each other within the experimental errors. 
The data may be represented by a single average 
cross section value of 
$\der\sigma(W,z)/\der z = (8.05 \pm 0.06 \stat  \pm 0.89 \syst) \mub$.
The restriction in the transverse momentum of the final state proton to 
$\ptproton \leq 0.2\gev$
implies that the measured cross section represents 
$(23 \pm 2 \stat \pm 5 \syst )\%$ of the
full differential cross section $\der\sigma/\der z$ in  
photoproduction if the $t$~dependence is assumed to follow~$e^{bt}$ 
with~$b=(6.6\pm 0.7\stat\pm 1.5\syst)\gev^{-2}$  
(see section~\ref{secregge}). 
The ratio of photoproduction
events with a leading proton of $0.66 \leq z \leq 0.90$ and
$p_T<0.2\gev$ to all photoproduction events~\cite{h1gammap} 
is $(1.17 \pm 0.02\stat  \pm 0.15 \syst )\%$.

From the observation that the cross section $\der \sigma/ \der z$ 
is independent of $z$ and $W$ within errors, we
infer that $\der \sigma/ \der M_X^2$ is also approximately
independent of $M_X$ at fixed $W$.
This observation is different from 
pomeron exchange processes, dominant at lower~$M_X/W$, 
where a $\der \sigma/ \der M_X^2 \sim 1/M_X^2$ 
dependence is a good approximation to the data\cite{H1dd}.
The present measurement is reminiscent of 
data from proton-proton scattering~$pp \to Xp'$
at lower centre-of-mass energies squared 
$s < 3900 \gev^2$~\cite{albrow,Goulianos}, where a flattening
of the cross section $\der \sigma/ \der M_X^2$ is observed for 
masses above $M_X/\sqrt s \sim 0.2$.

\vspace{0.5cm}

In section~\ref{secregge}, the measured cross-section 
$\der\sigma / \der z$ is interpreted in a Regge motivated
model. The lack of $W$~dependence of the cross section is also
supportive of the hypothesis of limiting fragmentation~\cite{limitfrag}
which states that target fragmentation is independent of projectile
energy and type.

\subsection{Triple Regge Analysis}
\label{secregge}

\begin{figure}[ht]
\begin{center}
\epsfig{file=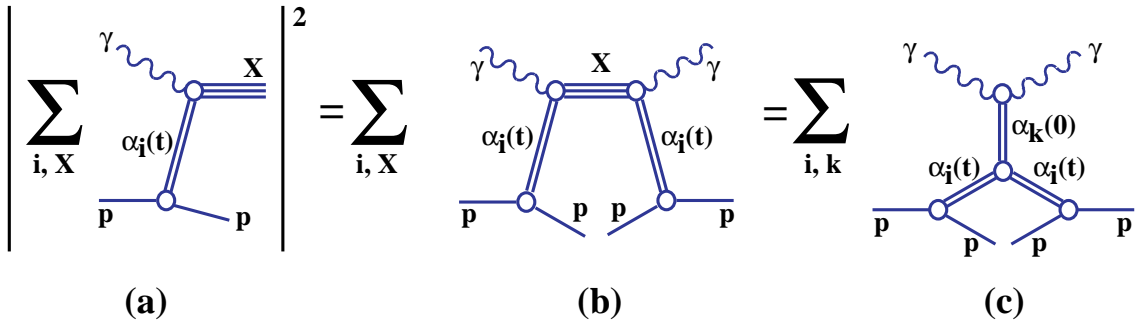,width=15cm}
\caption{ \sl 
Illustration of the Mueller-Regge model for the inclusive photon dissociation
cross section.
}
\label{phdd}
\end{center}
\end{figure}

In the language of Regge phenomenology, the~$W$ and $z$~dependences
of $\der\sigma/\der z$ yield
information about the exchange mechanisms contributing to 
leading proton photoproduction. 
A framework to model the process $\gamma p \to Xp'$ 
is offered by ``Triple Regge'' phenomenology using the Mueller-Regge 
approach\footnote{This approach is normally recommended for 
the kinematic range $1-z = M_X^2/W^2 \ll 1$. 
However, since the exact region of applicability is uncertain, 
the approach has been 
applied here to test the validity of the model and to investigate the 
contributing exchanges.}~\cite{AlM}, as illustrated in figure~\ref{phdd}. 
The total cross section at fixed $M_X$ is obtained from a coherent sum of the 
amplitudes for the exchange of Regge trajectories~$\alpha_i$, 
figure~\ref{phdd}a, which is related, through the generalised optical 
theorem \cite{AlM}, to the forward amplitude (figure \ref{phdd}b) for 
the process $\gamma \alpha_i(t) \rightarrow \gamma \alpha_i(t)$ at an 
effective centre of mass energy $M_X$. 
If $M_X^2$ is much larger than the hadronic mass scale $s_0$ 
and $W^2 \gg M_X^2$, this forward amplitude 
can be expressed as a further sum of 
Regge trajectories~$\alpha_k$ as shown in figure~\ref{phdd}c. 
Here we neglect the 
possibility of interference contributions, such that the two Regge 
trajectories coupling to the proton in the Triple Regge diagram of 
figure~\ref{phdd}c are always identical. 
The cross section can 
then be expressed as a sum over the contributing Regge trajectories~\cite{8-11}
\begin{equation}
\label{regge}
\frac{\der^2\sigma}{\der t\der z}=W^2 \frac{\der^2\sigma}{\der t \der M_X^2}
=\frac{s_0}{W^2}\sum_{i,k} 
G_{iik}^{\gamma p \rightarrow X p}(t) \left(\frac{W^2}{M_X^2}\right)
^{2\alpha_i(t)}
\left(\frac{M_X^2}{s_0}\right)^{\alpha_k(0)} \ ,
\end{equation}
where $\alpha_i$ refers to the trajectory exchanged between the photon and
the proton and $\alpha_k$ refers to the additional trajectory describing the 
total cross section between~$\alpha_i$ and the photon.
The product of the couplings of trajectory $i$ to the proton, 
trajectory $k$ to the photon and the three-Reggeon ($iik$) coupling 
is represented by $G_{iik}^{\gamma p \rightarrow X p}(t)$. 

In \cite{H1dd}, Triple Regge 
fits to lower $M_X$ photoproduction data showed dominant  
pomeron exchange at the largest $z$ and $W$, with further trajectories 
consistent with $\reg$ contributing at smaller $W$ or $z$. 
At the larger $M_X/W$ values of the present study, the full mixture of 
contributing trajectories and their interferences is presumably rather 
complicated. The data are not sufficiently precise and do not cover a 
wide enough kinematic region to make a full decomposition. Instead
we consider single effective trajectories $\alpha_i$ and 
$\alpha_k$ and assume 
that these represent averages of the mixtures contributing to the reaction. 
It is further assumed that all $t$~dependent terms can be absorbed 
into a single exponential~$e^{bt}$, 
where $b=b_0 - 2 \alpha_i^\prime \ln(1-z)$.
The $z$~dependent term in this expression
arises from the $t$~dependence of the trajectory
of the form $\alpha_i(t)=\alpha_i(0)+\alpha_i^\prime t$.
With these assumptions, after integration over~$t$, equation~\ref{regge} 
becomes 
\begin{equation}
\label{reggesim}
\frac{\der\sigma}{\der z}=W^2 \frac{\der\sigma}{\der M_X^2}=
\frac{A}{b}
 \ s_0^{1-\alpha_k(0)} \ (e^{bt_{min}}-e^{bt_{max}})(W^2)^{2\alpha_i(0)-1}
(M_X^2)^{\alpha_k(0)-2\alpha_i(0)}
\end{equation}
where $|t_{min}|$ is the minimum kinematically allowed value of $|t|$,  
$|t_{max}|$ is the maximum value allowed by the experimental limit of 
$p_T \le 0.2\gev$ for the leading proton, $A$ is an overall 
normalisation factor and $s_0$ is taken to be $1\gev^2$ following the
usual convention~\cite{Goulianos}. 

\begin{figure}[p]
    \begin{center}
        \setlength{\unitlength}{1mm}
        \epsfig{file=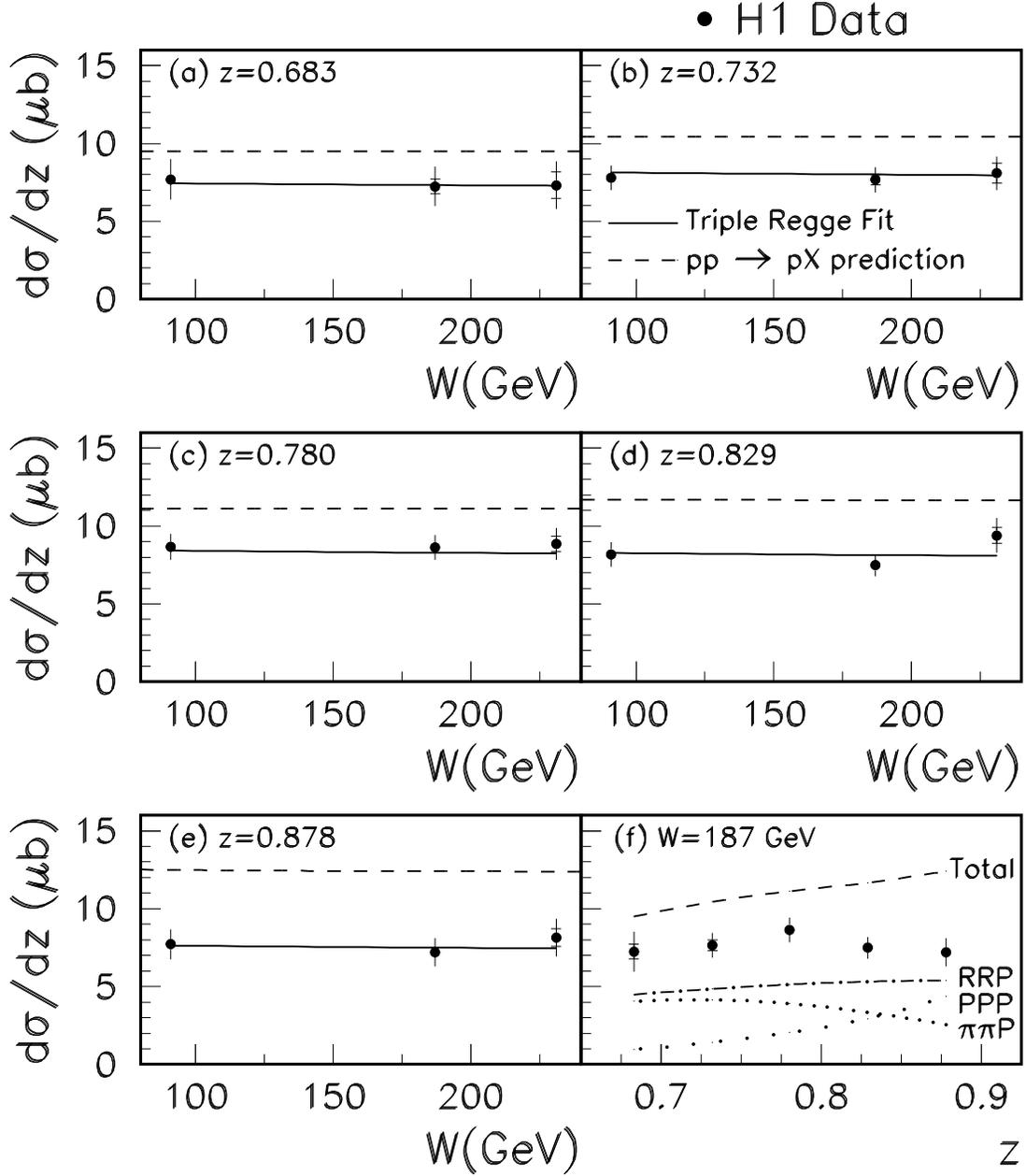, width=19cm}
        \caption[]{
        \sl (a-e) The differential cross sections $\der\sigma/\der z$ as
        a function of~$W$ for each measured $z$ value. The data are
        compared with the result of the Triple Regge fit
        described in the text (solid curves) and with the prediction
        derived from Triple Regge analysis of $pp \rightarrow X p^\prime$ 
        data \cite{GandR} (dashed curves). (f) The measured cross
        section as a function of $z$ at $W = 187 \ {\rm GeV}$, 
        compared with the prediction
        derived from Triple Regge analysis of $pp \rightarrow X p^\prime$ 
        data \cite{GandR}. The decomposition into the
        three dominant terms is also shown. 
        \label{picreggefit}
        }
    \end{center}
\end{figure}

The observed approximate independence of $\der\sigma/\der z$ of $M_X^2$ 
and $W$ 
implies in equation~\ref{reggesim} that $2\alpha_i(0)-1 \sim 0$ 
and $\alpha_k(0)-2\alpha_i(0) \sim 0$,
i.e.~$\alpha_i(0) \sim 0.5$ and $\alpha_k(0) \sim 1$. 
A fit of the full ansatz of equation~\ref{reggesim} to the data 
with $A$, $b_0$, $\alpha_i(0)$ and $\alpha_k(0)$ as free parameters,
taking $\alpha_i^\prime$ to be 
$(1.0 \pm 0.2)\gev^{-2}$~\cite{alphaprimerange,Goulianos}, gives
\begin{center}
\begin{tabular}{c @{$=$} c @{$\pm$} c @{$\stat $} c @{$\syst$} l}
$\alpha_i(0)$ & $0.33$ & $0.04$ & $\pm 0.04$ & , \\
$\alpha_k(0)$ & $0.99$ & $0.01$ & $\pm 0.05$ & , \\
$b_0$         & $(3.6$ & $0.7$  & $\pm 1.4 $ & $)\gev^{-2} \mbox{ and }$ \\ 
$A$           & $(495$ & $92$  & $^{+320}_{-205}$ & $)\mub/\gev^{2}$. 
\end{tabular}
\end{center}
In each case, the
first error quoted is the statistical and the second is the systematic 
uncertainty, which includes a contribution from the 
variation\footnote{Setting 
$\alpha_i^\prime$ to zero, the fit gives the almost 
identical values of $\alpha_i(0)=0.35$ and $\alpha_k(0)=0.99$ with similar 
errors.}
of $\alpha_i^\prime$. 
The fit gives a good representation of the data within the dominant
systematic errors. 
This is illustrated in figures~\ref{picreggefit}a-e which show the values 
of~$\der\sigma / \der z$ as a function of~$W$ in each $z$~bin with the fit,
represented by the solid curves, superimposed.


For the mean value in the data of~$z=0.78$, the value of~$b_0$
extracted from the fit implies 
that $b=(6.6 \pm 0.7\stat \pm 1.5 \syst )\gev^{-2}$.
The value of $b$ determined in this indirect manner
is compatible 
within experimental errors with that found by direct measurement
by ZEUS~\cite{ZEUSgamp} for $p_T^2 < 0.5 \ {\rm GeV^2}$.
The values of $\alpha(0)$ for the $\pi$, $\reg$ 
and Pomeron trajectories are approximately $0$, $0.5$ and $1.0$, 
respectively~\cite{Goulianos}. Hence, 
within the framework of the applied model,
we deduce from the extracted value of $\alpha_i(0)$ that the 
exchange between the photon and the proton can be understood as a mixture 
of trajectories, dominated by $\pi$ and~$\reg$, 
consistent with the previous H1 measurement 
at higher $Q^2$ \cite{lbproduction}. In addition since $\alpha_k(0)$ is 
close to unity, 
the total cross section for the scattering between the mixture 
of Regge trajectories
represented by $\alpha_i$ and the photon seems to be dominated by 
pomeron exchange, as would be expected for total cross sections at 
these energies ($32\gev < M_X < 130 \gev$).  

Inclusive proton reactions $pp \rightarrow X p^\prime$ have been
studied extensively in the framework of Triple Regge 
models \cite{RandR,FandF,GandR}. The present measurements are
compared with $pp \rightarrow X p^\prime$ data using the fits
described in \cite{GandR}. Vertex factorisation is assumed, such that
the ratio of couplings in the $\gamma p$ and
$pp$ cases is given by
\begin{equation}
  \left[ \
  \frac{G_{ii \pom}^{\gamma p \rightarrow X p} (t)}
       {G_{ii \pom}^{p p \rightarrow X p} (t)} \
  \right]
  =
  \left[ \
  \frac{\sigma_{tot}^{\gamma p}}
       {\sigma_{tot}^{p p}} \
  \right]_{s \rightarrow \infty} = 0.0031 \ ,
\end{equation}
where $s$ is the square of the centre of mass energy and the
numerical value of 0.0031 for the ratio of total cross sections
is taken from  \cite{dola}.
Equation~\ref{regge} is then used to 
obtain the $W$ and $z$ dependences of the cross section 
for $\gamma p \rightarrow X p^\prime$ 
assuming the couplings and
Regge trajectories given in \cite{GandR}.\footnote{The couplings
in table 3 of~\cite{GandR} from the fit with the restriction
$G^{p p \rightarrow X p}_{\pi \pi \pom} < 300 \ {\rm mb \ GeV^{-2}}$ 
are used. The
trajectories are $\alpha_{\pom}(t) = 1 + 0.25 \ t$,
$\alpha_{\reg}(t) = 0.5 + t$ and $\alpha_\pi(t) = t$. The $t$
dependence of each Triple Regge coupling is parameterised with
a single exponential.} In this model,
the dominant contributions for the present kinematic range are
$\pi\pi\pom$, $\reg\reg\pom$ and $\pom\pom\pom$, all other
terms being negligible. The total predicted cross section is
shown in figures~\ref{picreggefit}a-e (dashed curves). The  
prediction at $W = 187 \ {\rm GeV}$ and its decomposition into
the three dominant Triple Regge terms are shown in 
figure~\ref{picreggefit}f.


The predicted photoproduction cross sections lie
$25 - 65\%$ above
the measured values (see figure~\ref{picreggefit}), which
can be considered as fair agreement given the simplicity of the
model and the uncertainties. 
Hence $pp \rightarrow Xp^\prime$ and
$\gamma p \rightarrow X p^\prime$ data away from the pomeron
exchange dominated region can be reasonably well described 
within a single relatively simple Triple Regge model.
This contrasts with the discrepancies of an order of magnitude or more
observed when predicting diffractive hard scattering cross sections 
at larger $z$ in $\bar{p}p$ collisions at the Tevatron~\cite{Hannes}
using parton densities 
extracted from diffractive DIS at HERA~\cite{H1f2d3,ZEUSf2d3}.

\subsection{Leading Proton Production in DIS and Photoproduction}
\label{secgammapanddis}
In this section, the photoproduction cross sections with a leading proton
are combined with 
data from deep inelastic scattering with a leading proton in the 
range $0.71 < z < 0.90$, published in~\cite{lbproduction}. 
Since the DIS and photoproduction leading proton
data were recorded under very similar
experimental conditions, the bulk of the systematic errors
are identical and thus have little impact on the analysis.
Comparing the photoproduction and DIS data allows one to study changes in the 
scattering process as the incoming projectile changes from a hadron-like object
at~$Q^2=0$ to a point-like probe at higher~$Q^2$.
The DIS data sample covered the range 
$2\gev^2 \leq Q^2 \leq 50\gev^2$
and $6\cdot 10^{-5} \leq x \leq 6\cdot 10^{-3}$, 
where~$x$ is the Bjorken scaling variable.
A leading proton structure function $\ftwolpthree$ was defined as:
\begin{equation}
\label{formf2lpthree}
\frac{\der^3\sigma}{\der x \der Q^2 \der z} = 
    \frac{4\pi\alpha^2}{xQ^4}
    \left( 1-y+\frac{y^2}{2[1+R(x,Q^2,z)]} \right) \ftwolpthree(x,Q^2,z),
\end{equation}
where $\alpha$ denotes the electromagnetic coupling constant,
$y$ the inelasticity variable and
$R(x,Q^2,z)$ the ratio of the longitudinal to the
transverse polarised photon induced
DIS cross sections with a leading proton. 
Due to the $y^2$~factor, the impact of 
$R(x,Q^2,z)$ on the measurement of the structure function 
$\ftwolpthree(x,Q^2,z)$ is small
and $R$ was set to zero in~\cite{lbproduction}. 
The structure function $\ftwolpthree$ is related to the 
$\gamma^* p$~cross section by
\begin{equation}
\label{formlpsigmatot}
\frac{\der\sigma_{\gamma^{(*)} p \to Xp'}(Q^2,W,z)}{\der z} =
    \frac{4\pi^2\alpha}{Q^2}\ftwolpthree(x,Q^2,z).
\end{equation}

\begin{figure}[ht]
    \begin{center}
        \setlength{\unitlength}{1mm}
        \epsfig{file=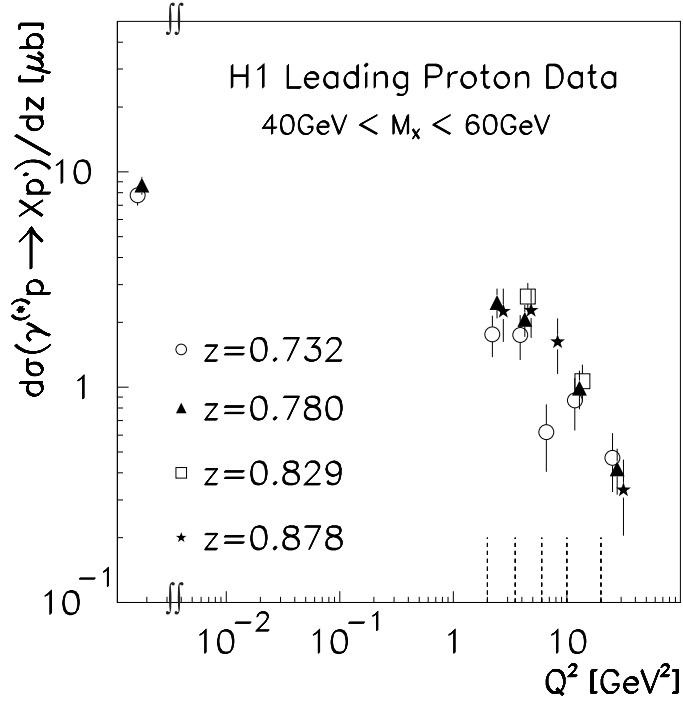,width=10cm}
        \caption[]{
  \sl The cross section $\der \sigma_{\gamma^{(*)}p\to Xp'}/\der z$ 
      as a function of $Q^2$ for
      four values of $z$ in the interval 
      $40\gev <  M_X < 60\gev $. The DIS data points are derived
from~\cite{lbproduction}. The data points 
   in the same $Q^2$ bin are displaced in $Q^2$ for visibility.
   The dashed lines indicate the bin limits in~$Q^2$.
        \label{plotsigmatotlp}
        } 
    \end{center}
\end{figure}

The DIS cross sections $\der\sigma/\der z$ deduced from the values of
$\ftwolpthree$ using equation~\ref{formlpsigmatot}, are shown together with the
photoproduction measurement reported here in figure~\ref{plotsigmatotlp}
for the data in the range $40\gev < M_X < 60\gev$ at
$z=0.732$, $0.780$, $0.829$ and $0.878$. The data at other values of~$M_X$
(not shown) have a very similar $Q^2$~dependence.
For all values of~$z$ investigated, the data approach $Q^2=0\gev^2$ in
a qualitatively similar manner to that observed for the inclusive
$ep$~cross section~\cite{h1f2fits,zeuslowq2}.

A more sensitive quantity with which to study the transition from DIS to
photoproduction is the ratio
of the semi-inclusive to inclusive cross sections.
For photoproduction, the ratio~$\fraclp$ of leading proton production 
per unit~$z$ to the total cross section is defined as
\begin{equation}
\fraclp (Q^2\approx 0,W,z) = 
 \frac{\der\sigma_{\gamma p \to Xp'}/\der z (Q^2\approx 0,W, z)}%
      {\sigma_{tot}^{\gamma p}(W)},
\end{equation}
where $\sigma_{tot}^{\gamma p}$ was taken from the Donnachie and Landshoff 
parametrisation in~\cite{dola} of the form
\begin{equation}
\label{dola}
\sigma_{tot}^{\gamma p} \ = \
68 \ \left( \frac{W^2}{\rm GeV^2} \right)^{0.0808} \ + \ 
129 \ \left( \frac{W^2}{\rm GeV^2} \right)^{-0.4525} \ {\rm \mu b}.
\end{equation}
For DIS, the ratio of leading proton 
production per unit~$z$ to the inclusive process is obtained from 
\begin{equation}
\fraclp (Q^2,W,z)
= \frac{\ftwolpthree(x,Q^2,z)}{F_2^{\mathrm{H1QCD}}(x,Q^2)},
\end{equation}
where $F_2^{\mathrm{H1QCD}}(x,Q^2)$ is a parametrisation of the proton
structure function taken from a QCD fit to the data in~\cite{h1f2}. Note
that $W\approx\sqrt{Q^2/x}$ at low~$x$. 

\begin{figure}[ht]
    \begin{center}
        \setlength{\unitlength}{1mm}
        \epsfig{file=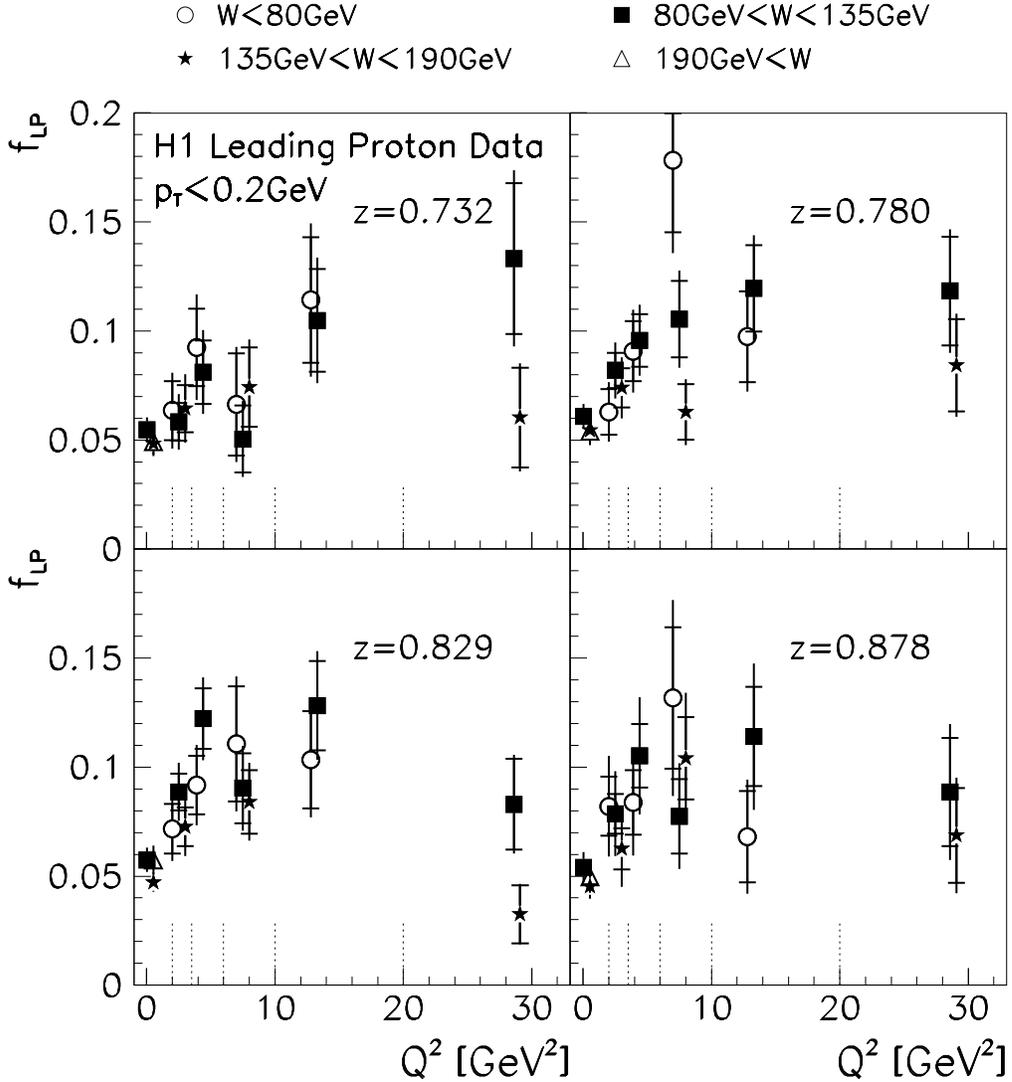,width=15cm}
        \caption[]{
  \sl The ratio per unit~$z$, $\fraclp$, of the number of leading proton
  events for four values of $z$ to the total in four intervals of $W$ as a
  function of $Q^2$.  The data points in the same $Q^2$ bin are
  displaced in~$Q^2$ for visibility.  The dashed lines indicate the bin
  limits in $Q^2$.
        \label{plotratios}
        } 
    \end{center}
\end{figure}

The ratios $\fraclp(Q^2,W,z)$ are shown in figure~\ref{plotratios},
separately for each 
$z$~value in four ranges of~$W$. 
The fraction of leading proton events in the $z$~range under study
is found to increase\footnote{
The $\chi^2$ for the DIS data in figure~\ref{plotratios} to be 
compatible with the mean values of the photoproduction data
is computed for each $z$~value. The result is a 
total~$\chi^2$ of $141$ for $48$~degrees of freedom, evaluated using
the full experimental errors.}
with~$Q^2$ from $\sim 5\%$ per unit~$z$ in photoproduction 
($Q^2\sim 0\gev^2$) 
to about $10\%$ per unit~$z$ at $Q^2 \sim 10\gev^2$ at all $z$ 
and~$W$.

The observed rise of $\fraclp$ from low to high~$Q^2$ in 
figure~\ref{plotratios} cannot fully
be explained by any reasonable $Q^2$~dependence of the $t$~slope
and thus indicates a breaking of factorisation of
the proton from the photon vertex.
A similar suppression of the cross section relative to predictions based
on vertex factorisation has been
observed in diffractive dijet production in regions where the photon is
resolved~\cite{h1jets} and in the process $pp \rightarrow Xp$ at high
energy~\cite{diffrinpbarp}.
Absorptive corrections, corresponding to multiple
Reggeon exchange, would be expected to give rise to such an 
effect~\cite{abscorr}.
It could also be
due to the onset with increasing $Q^2$ of the cross section for 
longitudinally polarised photons if
the relative contribution of the longitudinal cross section is 
different in the semi-inclusive and inclusive cases.

\section{Summary}
Photoproduction reactions with a final state proton of $\ptproton\leq 0.2 \gev$
observed in the H1 Forward Proton Spectrometer (FPS) have
been analysed in the kinematic range $0.66 \leq z \leq 0.90$
at $\gamma p$~centre-of-mass energies $W=91$, $187$ and $231\gev$. The 
cross section $\der \sigma(W,z)_{\gamma p \to Xp'} /\der z$ was determined to
be on  average $8.05 \pm 0.06 \stat  \pm 0.89  \syst \mub$, 
independent of $z$ and $W$ within the experimental errors.

A fit of a Triple Regge model to the data gives a good representation.
In this model,
the proton-photon interaction can be described
by the exchange of an effective trajectory having
$\alpha_i(0)=0.33 \pm 0.04 \stat  \pm 0.04  \syst $,
consistent with a mixture of pions and members of the exchange
degenerate trajectory $\reg$ with $\alpha_{\reg}(0) \sim 0.5$.
This result is consistent with the findings at similar~$z$ in deep inelastic 
scattering.
The total cross section for the interaction of the mixture of
trajectories $\alpha_i$ with the photon is described
by $\alpha_k(0)=0.99 \pm 0.01\stat  \pm 0.05 \syst $,
implying that it is dominated by the pomeron term.
The exponential slope parameter describing the $t$~dependence
of the data is indirectly determined within this model to be
$b=(6.6 \pm 0.7 \stat \pm 1.5\syst)\gev^{-2}$ at $z = 0.78$. 
A more detailed Triple Regge model of $pp \rightarrow Xp$ data~\cite{GandR}
has been extended to describe dissociative photoproduction by assuming 
vertex factorisation and gives a reasonable description of the measurements.

The data from photoproduction and
deep inelastic scattering with a final state proton
observed in the FPS
show a qualitatively similar $Q^2$~dependence to that observed when 
going from virtual to real photons in the inclusive reaction
$\gamma^{(*)} p \to X$.
The cross section for leading proton production 
expressed as a fraction per unit $z$ 
of the total $\gamma^{(*)}p$~cross section,
has been studied 
as a function of $Q^2$ in intervals of $z$ and $W$.  
This fraction has been found to increase with~$Q^2$.

\section*{Acknowledgements}

We are grateful to the HERA machine group whose outstanding
efforts have made and continue to make this experiment possible. 
We thank
the engineers and technicians for their work in constructing and now
maintaining the H1 detector, our funding agencies for 
financial support, the
DESY technical staff for continual assistance 
and the DESY directorate for the
hospitality which they extend to the non DESY 
members of the collaboration.
The forward proton spectrometer was supported by the 
INTAS93-43 project and the NATO contract PST.CLG.975100.


\begin{table}
\begin{center}
\include{tabtex}
\caption[]{\sl 
The differential cross section $\der\sigma_{\gamma p \to Xp'}/ \der z(W,z)$  
in $\mub$
for proton transvere momenta $p_T < 0.2\gev$
for five intervals in~$z$ and three intervals in~$W$ in $\gev$.
The uncertainties quoted are the contributions arising from 
the statistical error (\,{\sf stat}), 
the uncertainty introduced by the $z$ measurement (\,{\sf fps}),
errors depending only on the $W$ interval (\,{\sf wbin}), 
and a common normalisation uncertainty (\,{\sf norm}).
The total relative error is given in brackets.
\label{tabxsect}
}
\end{center}
\end{table}

\end{document}

%% file: h1auts.tex

C.~Adloff$^{33}$,              
V.~Andreev$^{24}$,             
B.~Andrieu$^{27}$,             
T.~Anthonis$^{4}$,             
V.~Arkadov$^{35}$,             
A.~Astvatsatourov$^{35}$,      
I.~Ayyaz$^{28}$,               
A.~Babaev$^{23}$,              
J.~B\"ahr$^{35}$,              
P.~Baranov$^{24}$
% ...

\bigskip{\it
% ...
$ ^{4}$ Inter-University Institute for High Energies ULB-VUB, Brussels;
          Universitaire Instelling Antwerpen, Wilrijk; Belgium$^{ c}$ \\
% ...
$ ^{23}$ Institute for Theoretical and Experimental Physics,
          Moscow, Russia \\
 $ ^{24}$ Lebedev Physical Institute, Moscow, Russia$^{ e,h}$ \\
% ...
$ ^{27}$ LPNHE, Ecole Polytechnique, IN2P3-CNRS, Palaiseau, France \\
 $ ^{28}$ LPNHE, Universit\'{e}s Paris VI and VII, IN2P3-CNRS,
          Paris, France \\
% ...
$ ^{33}$ Fachbereich Physik, Bergische Universit\"at Gesamthochschule
          Wuppertal, Wuppertal, Germany$^{ a}$ \\
% ...
$ ^{35}$ DESY, Zeuthen, Germany$^{ a}$ \\
% ...
$ ^a$ Supported by the Bundesministerium f\"ur Bildung, Wissenschaft,
      Forschung und Technologie, FRG,
      under contract numbers 7AC17P, 7AC47P, 7DO55P, 7HH17I, 7HH27P,
      7HD17P, 7HD27P, 7KI17I, 6MP17I and 7WT87P \\
% ...
$ ^c$ Supported by FNRS-NFWO, IISN-IIKW \\
% ...
$ ^e$ Supported by the Deutsche Forschungsgemeinschaft \\
% ...
$ ^h$ Supported by Russian Foundation for Basic Researc
      grant no. 96-02-00019 \\
% ...
}

%% file: tabtex.tex
 \begin{tabular}{|c|c|c@{\quad$\pm$}c@{$\pm$}c@{$\pm$}c@{$\pm$}c@{\quad($\pm$}r@{) }|}
 \hline
 \hline
 \rule[-2.5mm]{0mm}{7mm} $z$ & $\langle W_{\gamma p} \rangle$ &$\der\sigma/\der z$ &  {\sf stat} & {\sf fps} &
 {\sf wbin} & {\sf norm}   & {\sf total} \\
\rule[-2.5mm]{0mm}{6mm}      &$[\gev]$&  \multicolumn{5}{l}{$[\mub]$} &
\multicolumn{1}{c|}{ }       \\   
\hline
 \hline
0.683 &   91 &   7.69 & 0.22 &  1.09 & 0.58 & 0.41  & 17.1\%\\
0.683 &  187 &   7.24 & 0.47 &  1.02 & 0.47 & 0.38  & 17.7\%\\
0.683 &  231 &   7.31 & 0.86 &  1.03 & 0.62 & 0.39  & 20.9\%\\
 \hline
0.732 &   91 &   7.79 & 0.16 &  0.34 & 0.59 & 0.41  & 10.4\%\\
0.732 &  187 &   7.66 & 0.33 &  0.33 & 0.50 & 0.41  & 10.4\%\\
0.732 &  231 &   8.09 & 0.63 &  0.35 & 0.69 & 0.43  & 13.4\%\\
 \hline
0.780 &   91 &   8.67 & 0.13 &  0.16 & 0.66 & 0.46  &  9.5\%\\
0.780 &  187 &   8.64 & 0.28 &  0.15 & 0.56 & 0.46  &  9.2\%\\
0.780 &  231 &   8.85 & 0.50 &  0.16 & 0.75 & 0.47  & 11.6\%\\
 \hline
0.829 &   91 &   8.19 & 0.12 &  0.21 & 0.62 & 0.43  &  9.7\%\\
0.829 &  187 &   7.49 & 0.25 &  0.19 & 0.49 & 0.40  &  9.4\%\\
0.829 &  231 &   9.40 & 0.51 &  0.24 & 0.80 & 0.50  & 11.7\%\\
 \hline
0.878 &   91 &   7.71 & 0.14 &  0.65 & 0.59 & 0.41  & 12.7\%\\
0.878 &  187 &   7.19 & 0.29 &  0.61 & 0.47 & 0.38  & 12.6\%\\
0.878 &  231 &   8.15 & 0.56 &  0.69 & 0.69 & 0.43  & 14.8\%\\
 \hline
 \hline
 \end{tabular}